\documentclass[aip,jcp,reprint]{revtex4}
\usepackage{amsmath,amsfonts,amssymb}
\usepackage[english]{babel} 
\usepackage[latin1]{inputenc} 
\usepackage[T1]{fontenc}
\usepackage{color}
\usepackage{float}
\usepackage{verbatim}
\usepackage{graphicx}
\usepackage{bm}
\usepackage{mathtools}

\newcommand{\ve}[1][K]{\mathbf{#1}}

\def \equi#1{\mathrel{\mathop{\kern 0pt\sim}\limits_{#1}}}

\begin{document}

\title{Non-Markovian closure kinetics of flexible polymers with hydrodynamic interactions}

\author{N. Levernier$^1$, M. Dolgushev$^{2}$, O. B\'enichou$^1$, A. Blumen$^{2}$, T. Gu\'erin$^{3}$, R. Voituriez$^{1,4}$}

\affiliation{$^{1}$Laboratoire de Physique Th\'eorique de la Mati\`ere Condens\'ee, CNRS/UPMC, 
 4 Place Jussieu, 75005 Paris, France}
\affiliation{$^{2}$Theoretical Polymer Physics, University of Freiburg, 
   Hermann-Herder-Str. 3, D-79104 Freiburg, Germany}
 \affiliation{$^{3}$Laboratoire Ondes et Mati\`ere d'Aquitaine, University of Bordeaux, Unit\'e Mixte de Recherche 5798, CNRS, F-33400 Talence, France }
\affiliation{$^{4}$Laboratoire Jean Perrin, CNRS/UPMC, 
 4 Place Jussieu, 75005 Paris, France}

\begin{abstract}
This paper presents a theoretical analysis of the closure kinetics of a polymer with hydrodynamic interactions. This analysis, which takes into account the non-Markovian dynamics of the end-to-end vector and  relies on the preaveraging of the mobility tensor (Zimm dynamics),  is shown to reproduce very accurately the results of numerical simulations of the complete non linear dynamics. It is found that Markovian treatments, based on a  Wilemski-Fixman approximation, significantly overestimate cyclization times (up to a factor 2), showing the importance of memory effects in the dynamics.  In addition, this analysis provides scaling laws of the mean first cyclization time (MFCT) with the polymer size $N$ and capture radius $b$, which are identical in both Markovian and non-Markovian approaches. In particular, it is found that the scaling of the MFCT for large $N$ is given by $T\sim N^{3/2}\ln (N/b^2)$, which differs from the case of the Rouse dynamics where $T\sim N^{2}$. The extension to the case of the  reaction kinetics of a  monomer of a Zimm polymer with an external target in a confined volume is also presented.

\end{abstract}

\maketitle

\section{Introduction}

Reactions involving polymers are ubiquitous in nature. Among them, reactions of closure of linear chains are of particular interest since they  are involved  in  a number of chemical and biological processes. Examples cover  gene regulation by the formation of RNA-hairpins\cite{Bonnet1998,Wang2004} or DNA-loops\cite{Wallace2001,Allemand2006},  the folding of polypeptides\cite{Lapidus2000,Moeglich2006}, as well as the appearance of cycles in synthetic polymers \cite{Gooden1998,Zheng2011,Burgath2000}. In the diffusion-controlled regime, the kinetics of contact formation strongly depends on the complex dynamics of the reactive monomers.  As a result of the collective dynamics of all the monomers in the chain, the motion of a single monomer  is often subdiffusive\cite{Edwards} and presents non-Markovian features\cite{Panja2010} (\textit{i.e.} memory effects), which lead to  nontrivial reaction kinetics\cite{DEGENNES1982} even for the simplest models of polymers. 

On the theoretical level, various approaches have been proposed to quantify the kinetics of polymer closure\cite{WF,WILEMSKI1974b,Szabo1980,Sokolov2003,Likthman2006,ThomasNat,Friedman1989,Chak,Amitai2012,Toan2008}. 
An important step has been provided by Wilemski and Fixman\cite{WF,WILEMSKI1974b}, who made a local equilibrium assumption, thereby replacing the non-Markovian problem by an effective Markovian approach. Other theories include the  so-called SSS analytical approach\cite{Szabo1980} 
, which neglects important aspects of the polymer dynamics, and the perturbative renormalization group theory\cite{Friedman1989}, which provides results at leading order in the parameter $\epsilon=4-d$, with $d$ the spatial dimension. Recent works have improved these approaches in the case of the Rouse chain (\textit{i.e.} flexible without hydrodynamic interactions) by introducing several methods: (i)  a refined way to take into account the memory of the initial configurations\cite{Sokolov2003}, (ii) an exact formal iterative resolution scheme in one dimension  \cite{Likthman2006}, (iii) a strong localized perturbation analysis \cite{Amitai2012}, and (iv) an approach  based on the calculation of the  distribution of chain configurations at the instant of cyclization\cite{ThomasNat,ThomasCyclization,ThomasTarget,Benichou:2015bh}, which has a strong influence on the contact kinetics. 
 
The Rouse model, however, provides an incorrect description of polymer chain dynamics in dilute solutions, where hydrodynamic interactions are long-ranged and deeply modify the motion of the polymer\cite{ZimmOriginal,Edwards} as well as the closure kinetics\cite{Friedman1989,Chak}. 
Since hydrodynamic interactions are non-linear, most of available analytical treatments involve approximations. A standard approach  consists in using a pre-averaged form introduced by Zimm\cite{ZimmOriginal}, which is known to be accurate  \cite{Edwards}.   In the context of closure kinetics, 
Ortiz-Repiso \textit{et al.}\cite{Repiso} found by simulations that the Wilemski-Fixman treatment of the cyclization kinetics of Zimm chains systematically overestimates the reaction times, in a way that cannot be attributed to the simplified treatment of these interactions. This suggests that non-Markovian effects, which  have not been described before in the presence of hydrodynamic interactions, play an important role in the cyclization kinetics. 

This paper is devoted to the theoretical description of the cyclization kinetics of chains with hydrodynamic interactions in the diffusion controlled regime. 
This problem has already been discussed in the Wilemski-Fixman approach\cite{Chak,Repiso} and the perturbative renormalization group theory\cite{Friedman1989}. However, a non-Markovian description of the closure kinetics has not been considered so far. The first goal consists in describing these effects by adapting a theory  proposed recently   \cite{ThomasNat,ThomasCyclization,ThomasSemiflex},  which  was so far restricted to the case of flexible and semi-flexible free draining chains. 
One could expect that the treatment of pre-averaged interactions is not valid for the problem of chain closure, since interactions  averaged  over equilibrium configurations could differ from the actual (non-preaveraged) interactions for closed chains,  which could then play an important role for the contact kinetics. We will show however that this effect is relatively small: indeed, calculating the average shape of the polymer at the instant of cyclization for  pre-averaged interactions yields a reaction time that is in quantitative agreement with the results of Brownian dynamic simulations performed  without pre-averaging.  
The second goal of the paper is to derive  new scaling laws for the mean cyclization time in the limit of  small capture radius and long chains. In particular, we prove that for long chains, the mean cyclization time $T$ scales as $T\sim N^{3/2}\log (N/b^2)$, with $N$ the number of monomers and $b$ the capture radius, for both the Wilemski-Fixman approach and the non-Markovian theory, with however a different prefactor. In addition, we predict scaling laws for intermolecular reactions, by extending  the formalism to the case of the search of an external target of size $r$ in a confined volume $V$. We show that in this case, the mean reaction time scales as $T\sim V \log (r/\sqrt{N})$ for long chains.

The outline of the paper is as follows. In section \ref{Zimm}, we recall briefly some important results of the  Zimm model. 
In section \ref{Non-mark}, we show how to adapt the recent non-Markovian theory  \cite{ThomasNat,ThomasCyclization,ThomasSemiflex} of cyclization kinetics to the Zimm  model. We compare the numerical solution of the model to both  Brownian dynamics simulations and to the results of a  Wilemski-Fixman approach. Section \ref{SectionScalings} is devoted to the derivation of scaling laws for the mean cyclization time with the chain length and the capture radius, for both the  Markovian  and the non-Markovian theories. Finally, we present in section \ref{External} the direct application of our calculations to the case of the reaction of one monomer with an external target in confinement and derive explicit asymptotic formulas for the reaction time for this problem.

\section{Polymer dynamics with hydrodynamic interactions}
\label{Zimm}

We consider the dynamics in a three-dimensional ($3d$) space of a polymer represented by $N$ beads at positions  $(\mathbf{x}_{1},...,\mathbf{x}_{N})$ (in this paper quantities in bold represent vectors or tensors in the $3d$ space). The beads are linked by springs of stiffness $k$, so that the force $\mathbf{F}_{j}$ exerted on a bead $j$ by the neighboring beads is 
\begin{align}
\mathbf{F}_{j}=-k(  2\ \mathbf{x}_{j}-\mathbf{x}_{j+1}-\mathbf{x}_{j-1})=-\sum_{k=1}^N M_{jk} \ve[x]_k, \label{Force}
\end{align}
where we used the convention $\ve[x]_0=\ve[x]_1$ and $\ve[x]_{N+1}=\ve[x]_N$ and $M$ is the (Laplacian) tridiagonal connectivity matrix. This force $\mathbf{F}_{j}$ is balanced by the force due to the fluid solvent, and from Newton's third law a force $\ve[F]_j$ is exerted on the fluid at the position of bead $j$. This force generates a velocity field in the fluid that influences the motion of all other beads. We introduce a non-isotropic mobility tensor $\ve[D]_{ij}$ to describe these hydrodynamic interactions, such that the average velocity of the fluid at the position of the bead $i$ is  $\ve[D]_{ij}\ve[F]_j$. 
The dynamics in the presence of hydrodynamic interactions then follows the Langevin equation in the  overdamped limit
\begin{equation}
	\dot{\mathbf{x}}_{i}=\sum_{j=1}^N \bold{D}_{ij}\cdot \mathbf{F}_{j} + \boldsymbol{\zeta}_i (t),
\label{Langevin}
\end{equation}
where $\boldsymbol{\zeta}_i(t)$ is a stochastic Gaussian white noise term whose amplitude follows from the fluctuation-dissipation relation, $\langle \boldsymbol{\zeta}_i(t)\otimes \boldsymbol{\zeta}_j(t') \rangle=2 k_BT\bold{D}_{ij} \delta (t-t')$, with $k_BT$ the thermal energy. We also introduce $l_0=\sqrt{k_BT/k}$ the typical bond length and $\tau_0=6\pi\eta a/k$ the single bond characteristic relaxation time, where $\eta$ is the fluid viscosity and $a$ the monomer radius. 

Different choices of mobility matrix $\ve[D]_{ij}$ exist. The simplest one is the Oseen tensor, describing the fluid velocity induced by a punctual force in the fluid, and therefore valid for quasi-punctual monomers. However, for very small distances between beads, the Oseen tensor may not be positive definite, making the dynamics  (\ref{Langevin}) unphysical due to the appearance of negative relaxation times \cite{BrownianDynamics}. Instead, for numerical purposes, we  use the so-called Rotne-Prager tensor, which  does not have this problem. We denote by  $\bold{r}_{ij}=\ve[x]_j-\ve[x]_i$ the vector from the particle $i$ to the particle $j$ and $r_ {ij}=\vert\bold{r}_{ij}\vert$. The elements of the Rotne-prager tensor are then given as follows\cite{RotnePragerOne}: when $i\ne j$, and when the beads $i,j$ do not overlap ($r_{ij}>2a$) , 
\begin{align}
&\ve[D]_{ij}=\nonumber\\
&\frac{k_{B}T}{8 \pi \eta  r_{ij}} \left[\left(1+\frac{2a^2}{3r_{ij}^2}\right) \bold{I}+\frac{\bold{r}_{ij}\otimes \bold{r}_{ij}}{r_{ij}^{2}} \left( 1-\frac{2a^{2}}{r_{ij}^{2}}\right) \right], \label{RP1}
\end{align}
while for $i\ne j$ and $r_{ij}<2a$,  
\begin{align}
\ve[D]_{ij}&=\frac{kT}{6 \pi \eta \,a} \left[\left( 1 - \frac{9}{32} \frac{r_{ij}}{a} \right) \bold{I}+ \frac{3}{32}\frac{\bold{r}_{ij} \otimes \bold{r}_{ij}}{r_{ij} \, a}  \right],\label{RP2}
\end{align}
and finally, when $i=j$ 
\begin{align}
	\ve[D]_{ii}&=\frac{k_{B}T}{6 \pi \eta \,a} \ve[I], \label{RP3}
\end{align}
where $\bold{I}$ is the $3\times 3$ identity tensor.  Eqs.~(\ref{Force},\ref{Langevin},\ref{RP1},\ref{RP2},\ref{RP3}) define the polymer  dynamics that will be analyzed in this paper. For numerical purposes, we will make  use of  the  choice $a=0.25\,l_0$ \cite{Inst,Inst2}. 

The non-linear dependence of the mobility matrix $\ve[D]_{ij}$ on the positions $\ve[x]_j$ makes the Langevin equation \eqref{Langevin} very difficult to solve. In particular it does not admit Gaussian solutions.  In his pioneering work\cite{ZimmOriginal}, Zimm overcame this difficulty by replacing the mobility matrix  $\ve[D]_{ij}$ by its average value over the equilibrium distribution, thereby making the equation linear. It has been shown that this approximation catches the main physics of hydrodynamic interactions \cite{ZimmOriginal,Edwards}. In the model introduced above, it would be therefore relevant to use a pre-averaged version of the Rotne-Prage tensor rather than the Oseen tensor.  However, it is shown in Appendix \ref{appPreav} that these tensors differ at most by a few percents for the parameters that we use. Hence, it is sufficient to use for analytical calculations the pre-averaged form of the  Oseen tensor, given by \cite{ZimmOriginal,Edwards}:
\begin{align}
\bar{\ve[D]}_{ij}\simeq \frac{k_BT}{6 \pi \eta} \left[\frac{\delta_{ij}}{a}+\frac{(1-\delta_{ij})}{l_0}\sqrt{\frac{2}{ \pi \vert i-j\vert}}\ \right]  \bold{I}.\label{PreaveragedTensor}
\end{align}
Note that the pre-averaged Oseen tensor is isotropic.  


The effects of hydrodynamic interactions on polymer dynamics have already been studied in details. In particular, it is well-known that any monomer of  a long polymer chain  performs a subdiffusive motion at intermediate time scales, which results from the collective dynamics of all the monomers.
More precisely, for $N \gg 1$ we have three different regimes for the mean-square displacement of a monomer \cite{Teraoka}~:
\begin{equation}
\left \langle [\bold{r}_{i}(t)-\bold{r}_{i}(0)]^2 \right \rangle \sim\left\{
\begin{aligned}
&D_{G} t& \; \;\text{for} \;\;& t \ll \tau_{0} \\
&A t^{2/3}&\; \;\text{for} \;\;& \tau_{0} \ll t \ll \tau_{0} N^{3/2} \\
&D_{G} t/\sqrt{N}& \; \;\text{for} \;\;& \tau_{0} N^{3/2} \ll t &
\end{aligned}
\right.
\label{psiMonom}
\end{equation}
where $D_{G}=k_{B}T/\eta l_{0}$, $A=(k_{B}T/\eta)^{2/3}$  (see Eq. (\ref{SubDIff})) and all the numerical constants have been omitted. We can already note that in the subdiffusion regime, the dimension of the random walk of a monomer (defined such that $\langle [r(t)-r(0)]^2\rangle\sim t^{2/d_w}$) is given by $d_w=3$. The process in $3d$ is therefore  marginally recurrent, whereas it is transient for both short and long time scales (see ref.\cite{Condamin2007}). This  will play  a key role in the determination of the scaling laws that  we present below.


The goal of this paper is to characterize the Mean First Cyclization Time (MFCT, denoted $T$), defined as the average first time for which the distance $\vert\ve[x]_N-\ve[x]_1\vert$ between the two chain ends becomes smaller than a capture radius $b$. Initially the polymer is in an equilibrium configuration, with the constraint that $\vert\ve[x]_N-\ve[x]_1\vert>b$. 
The effect of the complex monomer dynamics resulting from the hydrodynamic interactions will be analyzed with both numerical simulation methods (which we describe in App. \ref{num}) and analytical means (see section \ref{Non-mark}).





\section{Theories of cyclization  kinetics}
\label{Non-mark}

We present here a theoretical approach that enables the determination of the mean cyclization time $T$, which can be seen as the mean first passage time of a non-Markovian problem. The approach  consists in adapting a recent theory, developed so far for free-draining chains\cite{ThomasTarget,ThomasCyclization,ThomasSemiflex}, in which the MFCT is expressed in terms of the conformational distribution of chains at the very instant of cyclization, whose moments are computed by solving a set of self-consistent equations.

The main steps of this approach can be summarized as follows. 
We introduce the joint probability density  $f(\{\ve[x]\}, t)$ that the contact is made for the first time at time $t$ and that, at this first passage event, the macromolecule has a configuration described by the set of positions $\{\ve[x]\}=(\ve[x]_1,\ve[x]_2...)$. We partition the trajectories that lead to a configuration $\{\ve[x]\}$ (in which the contact condition is satisfied) into two steps, the first step consisting in reaching the target for the first time at $t'$, and the second step consisting in reaching the final configuration $\{\ve[x]\}$ in a time $t-t'$. The mathematical formulation of this decomposition of events is 
\begin{align}
	P(\{\ve[x]&\},t\vert \mathrm{ini},0)=\nonumber\\
&\int_0^t dt' \int d\{\ve[x]'\} f(\{\ve[x]'\},t') P(\{\ve[x]\},t-t'\vert \{\ve[x]'\}),\label{EquationRenewal}
\end{align}
where $d\{\ve[x]\}\equiv d\ve[x]_1d\ve[x]_2...d\ve[x]_N$, $P(\{\ve[x]\},t\vert \{\ve[x']\})$ is the probability of $\{\ve[x]\}$ at $t$ starting from $ \{\ve[x']\}$ at $t=0$ while $P(\{\ve[x]\},t\vert \mathrm{ini},0)$ is the probability of $\{\ve[x]\}$ at $t$ starting from the initial conditions at $t=0$ (in which the chain is at equilibrium, with the condition that the reactive monomers are not necessarily in contact).  
Next, taking the Laplace transform of (\ref{EquationRenewal}) and  expanding for small values of the Laplace variable, we obtain\cite{ThomasCyclization}
\begin{align}
T  &P_{\text{stat}} ( \{\bold{x}\} )= \nonumber\\
&\int_{0}^{\infty} dt  \left[ \int  d\Omega   \ P ( \{\bold{x}\},t | \pi_{\Omega})  -  P (\{ \bold{x}\},t |  \text{ini},0)\right].
\label{basis}
\end{align}
In this equation, $\Omega$ represents the angular directions parametrized by $\theta$ and $\varphi$ in spherical coordinates, with the normalization $\int d\Omega=1$; $\pi_{\Omega}(\{\ve[x]\})$ represents the probability distribution of configurations at the very instant of cyclization, given that the angular direction of the end-to-end vector at first contact is $\Omega$, and  $P ( \{\bold{x}\},t | \pi_{\Omega})$ is the probability of $\{\ve[x]\}$ at $t$ starting from the distribution $\pi_{\Omega}$ at initial time. Equation (\ref{basis}) is exact and does not depend on the particular hypotheses of chain dynamics. It is derived in details in Ref.\cite{ThomasCyclization}. However, this equation cannot be solved explicitly to the best of our knowledge  and approximations have to  be introduced.   

The first simplifying step is to approximate the dynamics by a Gaussian dynamics in order to be able to evaluate the propagators appearing in (\ref{basis}); hence we consider only the Zimm dynamics with the pre-averaged mobility tensor. Next, the simplest approach is to make a Markovian approximation, which consists in neglecting any memory effect by assuming that the distribution $\pi_{\Omega}$ is the equilibrium distribution conditional to   
 $\ve[x]_{N}-\ve[x]_{1}=b \hat{\bold{u}}_{r}(\Omega)$ (with $\hat{\bold{u}}_{r}(\Omega)$ the unit vector pointing in the direction $\Omega$). This corresponds to the so-called Wilemski-Fixman approximation\cite{Pastor,WF,ThomasCyclization}. Introducing this approximation into Eq. (\ref{EquationRenewal}), integrating over all configurations and taking the long time limit lead to the estimate $T_{\mathrm{WF}}$ of the MFCT \cite{ThomasCyclization}
\begin{align}
	T_{\text{WF}}=\int_0^{\infty} \frac{dt}{[1-\phi(t)^2]^{3/2}}\left\{ e^{-b^2\phi(t)^2/[2\psi(t)]}- \frac{Z(b,\psi(t))}{Z(b,L^2)}
\right\},\label{ExpressionTMarkovianCentre}
\end{align}
where $\phi(t),L^2,\psi(t)$ characterize the dynamics of the end-to-end vector $\ve[r]_{\mathrm{ee}}=\ve[x]_N-\ve[x]_1$, which is assumed to be Gaussian, and $Z(b,x)=\int_{b}^{\infty} dR_{0} e^{-R_{0}^{2}/2x}$. The function $\phi$ is the normalized temporal auto-correlation function of any coordinate of the end-to-end vector $\ve[r]_{\mathrm{ee}}$,
\begin{align}
\phi(\tau)=\frac{\langle x_{\mathrm{ee}}(t+\tau) x_{\mathrm{ee}}(t)\rangle}{\langle x_{\mathrm{ee}}(t) x_{\mathrm{ee}}(t)\rangle},\label{DefPhi}
\end{align}
where the spatial coordinates of $\ve[r]_{\mathrm{ee}}$ are $(x_{\mathrm{ee}},y_{\mathrm{ee}},z_{\mathrm{ee}})$, $L^2$ is the equilibrium mean-square length 
\begin{align}
L^2=\langle x_{\mathrm{ee}}(t)^2\rangle=(N-1)l_0^2,
\end{align}
and $\psi(t)$ is the Mean-Square-Displacement of any coordinate of $\ve[r]_{\mathrm{ee}}$ when the initial value of $\ve[r]_{\mathrm{ee}}$ is fixed, and is easily shown to be related to $\phi$ by
\begin{align}
	\psi(t)=\mathrm{Var}(x_{\mathrm{ee}}(t)\vert x_{\mathrm{ee}}(0) =x_{\mathrm{ee}}^0)
= L^2[1-\phi(t)^2], \label{DefPsi}
\end{align} 
where we denote $\mathrm{Var}(y\vert B)$ the variance of the variable $a$ given the event $B$ is realized.

\begin{figure}[h]
\begin{center}
\includegraphics[scale=0.40]{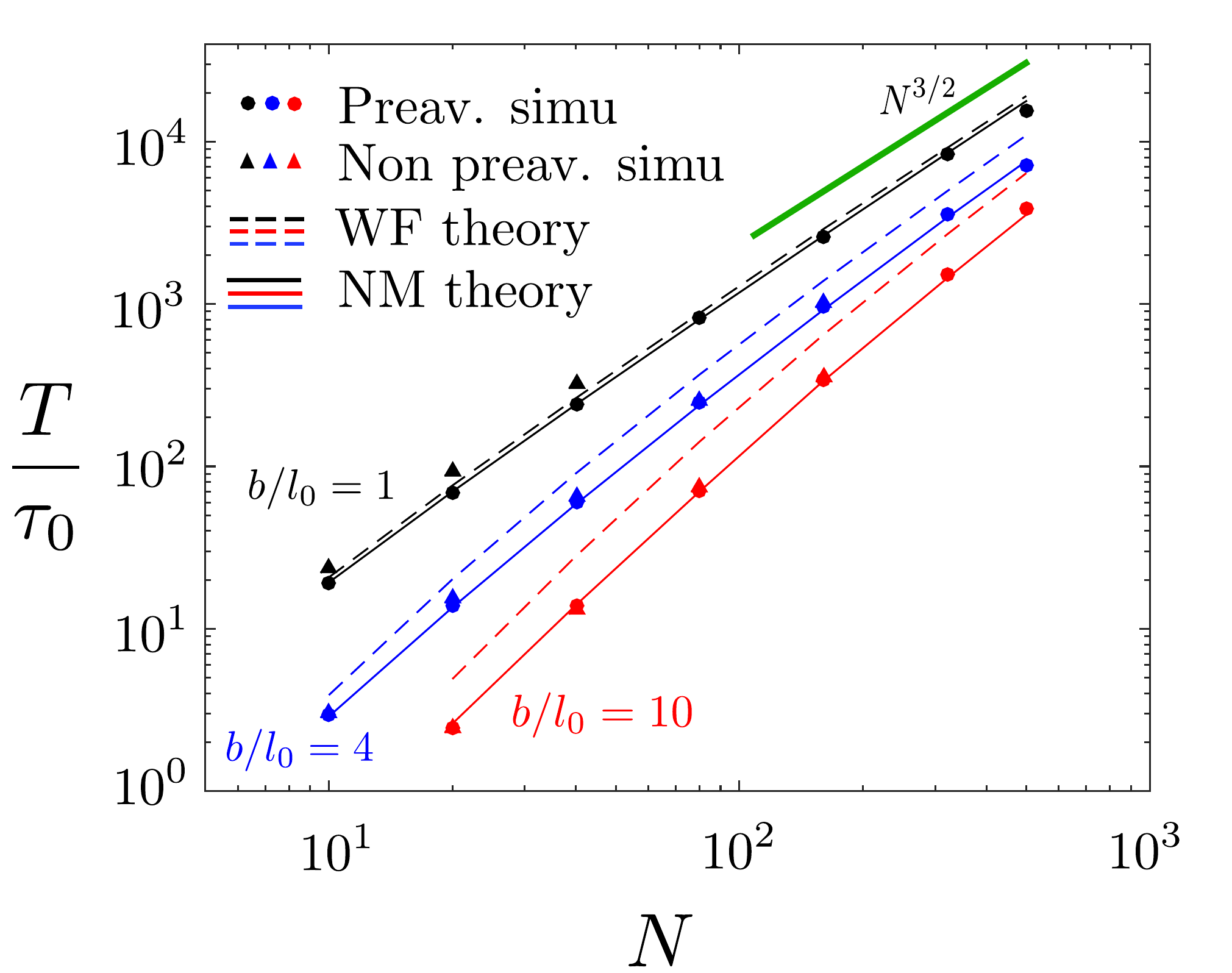} 
\end{center}
\caption{\textbf{Mean Cyclization Time of flexible chains with hydrodynamic interactions} 
MFCT as a function of the number of monomers $N$ for  different capture radii ($b=l_0$ in black, $b=4\,l_0$ in blue and $b=10\,l_0$ in red). Plain lines stand for the non-Markovian approach, dashed lines for  the Markovian approach. Dots stand for simulations with the pre-averaged Rotne-Prager tensor, and triangles for simulations without any pre-averaging. The asymptotic scaling $T\sim N^{3/2}$ obtained analytically (green thick line) is well reproduced. 
}
\label{timeFig}
\end{figure}

Going beyond the Wilemski-Fixman approximation requires a more precise description of the distribution $\pi_{\Omega}$. Here, our key hypothesis is to assume that $\pi_{\Omega}$ is a multivariate Gaussian distribution. 
For symmetry reasons, the average of $\bold{x}_{i}$ over $\pi_{\Omega}$ is along $\hat{\bold{u}}_{r}(\Omega)$. 
We denote this quantity $\mathbb{E}_{\pi_{\Omega}}(\ve[x]_{i})=m_{i}^{\pi} \hat{\bold{u}}_{r}(\Omega)$. 
We make the additional  assumption that the covariance matrix of $\pi_{\Omega}$ is the same covariance matrix that characterizes the equilibrium distribution of chains with the constraint $\ve[r]_{\mathrm{ee}}=b\hat{\bold{u}}_{r}(\Omega)$. Such an approximation is not necessary but largely simplifies the calculations. Previous studies on free-draining chains\cite{ThomasTarget,ThomasNat} have revealed that releasing this "stationary covariance approximation" only slightly improves the estimate of the cyclization time at an important calculation cost. 
A set of self-consistent equations that defines the $m_i^{\pi}$'s is found by multiplying (\ref{basis}) by $\ve[x]_i\,\delta(\ve[x]_N-\ve[x]_1-\ve[r]^*)$ (for a fixed $\ve[r]^*$ satisfying $\vert\ve[r]^*\vert<b$) and integrating over all configurations. Adapting existing calculations for free-draining chains\cite{ThomasCyclization}, the resulting equation is (see the complete derivation in Appendix \ref{appAutoCoh})
\begin{align}
\int_{0}^{\infty} \frac{dt}{\psi^{3/2}} \Bigg\{ & e^{-\frac{R_{\pi}^{2}}{2 \psi}}\left[\frac{R_{\pi}}{3 \psi}\left(\mu_{i}^{\pi}-\frac{\beta_{i}}{\psi} R_{\pi}\right)+\frac{\beta_{i}}{\psi}-\frac{\alpha_{i}}{L^{2}} \right] \nonumber \\
& + \left(\frac{\beta_{i}}{\psi}-\frac{\alpha_{i}}{L^{2}}\right) \frac{b^{3} e^{-b^{2}/2\psi}}{3Z(b,L^{2})}\Bigg\}=0 \label{mom}.
\end{align}
Here, we have denoted by $\mu_{i}^{\pi}(t)$ the average of $\ve[x]_i$ at time $t$ after the instant of  cyclization  in the direction $\hat{\ve[u]}(\Omega)$, $R_{\pi}=\mu_{N}^{\pi}-\mu_{1}^{\pi}$, $\beta_i(t)$ is the covariance between $x_i(t)$ and $x_{\mathrm{ee}}(t)$ when $x_{\mathrm{ee}}(0)$ is fixed, while $\alpha_i$ is the  covariance between $x_i$ and $x_{\mathrm{ee}}$ at equilibrium. 
In other words, $\alpha_i$ and $\beta_i$ characterize the dynamics and the equilibrium of the motion starting from a constrained equilibrium state, 
\begin{align}
&\beta_i(t)=\mathrm{Cov}(x_i(t),x_{\mathrm{ee}}(t)\vert x_{\mathrm{ee}}(0)=x_{\mathrm{ee}}^0),\label{DefBeta_i}\\
&\alpha_i=\mathrm{Cov}_{\mathrm{stat}}(x_i ,x_{\mathrm{ee}}), \label{DefAlpha_i}
\end{align}
(note that $\beta_i$ does not depend on $x_{\mathrm{ee}}^0$) and $\mu_{i}^{\pi},R_{\pi}$ characterize the motion of the chain in the future of the first contact,
\begin{align}
&\mu_{i}^{\pi}(t)=\left\langle \ve[x]_i(t+t^*)\cdot \frac{\ve[r]_{\mathrm{ee}}(t^*)}{\vert\ve[r]_{\mathrm{ee}}(t^*)\vert} \right\rangle,\\
&R_{\pi}(t)=\left\langle \ve[r]_{\mathrm{ee}}(t+t^*)\cdot \frac{\ve[r]_{\mathrm{ee}}(t^*)}{\vert\ve[r]_{\mathrm{ee}}(t^*)\vert} \right\rangle =\mu_{N}^{\pi}(t)-\mu_{1}^{\pi}(t),
\end{align}
where $t^*$ is the first cyclization time. The time evolution of $\mu_i^\pi$ follows from the Langevin equation
\eqref{Langevin}:
\begin{align}
	\partial_t\mu_{i}^{\pi}(t) =-\sum_{j,k=1}^N D_{ij}M_{jk}\mu_{k}^{\pi}(t), \hspace{0.5cm} \mu_{i}^{\pi}(0)=m_{i}^{\pi}. \label{095894}
\end{align}
Here, the unknowns $m_{i}^{\pi}$ are contained in the $\mu_{i}^{\pi}$ as initial conditions of the dynamical system (\ref{095894}), while $\alpha_i,\beta_i,\phi,\psi$ characterize the dynamics of the chain and are analytically calculated in Appendix \ref{appProp}. 
We stress that equation (\ref{mom}) is fully general for a $3d$ isotropic Gaussian non-Markovian process, and does not depend on the particular structure of the mobility matrix $D$ or the connectivity matrix $M$. 
Then, the expression of the mean reaction time is obtained by multiplying (\ref{basis}) by $\delta(\ve[x]_N-\ve[x]_1)$ and integrating over all configurations, leading to
\begin{equation}
\frac{T}{L^{3}}=\int_{0}^{\infty} \frac{dt}{\psi^{3/2}} \left[ \, e^{-R_{\pi}^{2}/2\psi}-\frac{Z(b,\psi)}{Z(b,L^{2})} \right].
\label{EquationMFCT}
\end{equation}
To conclude, Eq. \eqref{095894} provides a system of $N$ equations for the unknowns $\mu_{i}^{\pi}(t)$. Note that this system is of rank $N-2$, because the polymer center of mass can be set arbitrarily, and the constraint   $m_{N}^{\pi}-m_{1}^{\pi}=b$ must hold.   Solving this system allows us to compute the function $R_{\pi}(t)$, and then to calculate the mean first-passage time $T$ from \eqref{EquationMFCT}. 



We have solved numerically these equations for different values of the number of monomers $N$ and capture radius $b$. The results are shown in Fig. \ref{timeFig}. We clearly see that the non-Markovian theory accurately predicts  the mean-cyclization time, whereas the Wilemski-Fixman approach quickly fails when the capture radius is large enough. More precisely, the theory is in quantitative agreement with the simulations performed with preaveraging. Remarkably, the theory is also in quantitative agreement with the simulations performed with the exact (non pre-averaged) interactions, in which the full non-linear stochastic dynamics is taken into account when the capture radius is not small. 

\begin{figure}[h]
\begin{center}
\includegraphics[scale=0.45]{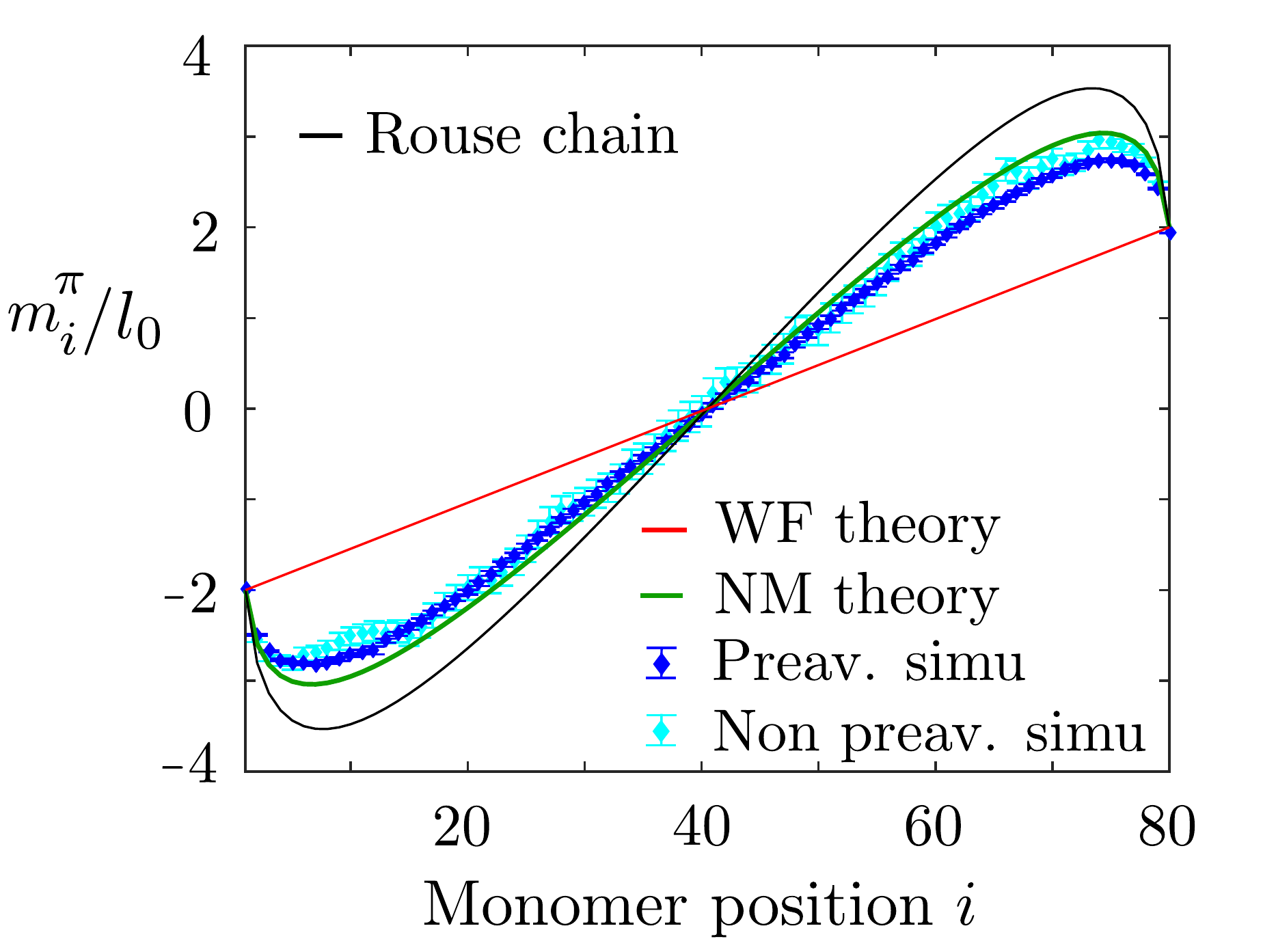} 
\end{center}
\caption{\textbf{Average position of the monomers $m_i^{\pi}$ at the instant of cyclization in the direction of the end-to-end vector at this instant.} We represent $m_{i}^{\pi}=\langle \ve[r]_i(t^*)\cdot\ve[u](t^*)\rangle$, with $t^*$ the first cyclization time and $\ve[u]=\ve[r]_{\mathrm{ee}}/\vert\ve[r]_{\mathrm{ee}}\vert$ for Zimm chains in the WF approximation (red line), the non-Markovian theory (green line), simulations with pre-averaging (blue diamonds) and without (diamonds in cyan). The result of the non-Markovian theory for Rouse chains is also shown for comparison (black line). Parameters $N$=80, $b=4\,l_0$. 
}
\label{splitting}
\end{figure}

The average positions $m_i^{\pi}$ of the monomers at the instant of cyclization are shown in Fig.~\ref{splitting}, which shows that the Zimm chain is significantly more elongated in the direction of the end-to-end vector at the reaction than in an equilibrium looped configuration. This allows the chain to perform cyclization more rapidly than if it had to equilibrate. The monomers  neighboring the reactive ones are on average outside the reactive region. Note that the small shift between the theoretical prediction and the simulation indicates that the theory is not exact. The shape of the curve of the $m_i^{\pi}$ is however significantly closer to the simulations for the non-Markovian theory than in the case of the Wilemski-Fixman approximation. 

For small capture radius, the cyclization times obtained numerically without pre-averaging  are slightly larger than those obtained numerically with pre-averaging, as well as those predicted by the non-Markovian theory. This can be understood from the fact that, for very small capture radius, the monomer motion at short time scales plays a key role. For small $t$, the MSD of the end-to-end distance reads $\psi(t) \simeq 4(D_{11}+D_{1N}) \, t$,  
and the motion is diffusive. With the pre-averaging, $D_{1N}\sim 1/\sqrt{N}$ is negligible compared to $D_{11}$. However,  for chains that are close to form a loop, the actual mobility tensor reads $D_{1N}\sim D_{11}$ since the distance between the end-beads is small. Hence, the effective diffusion constant at small time scales is not correctly estimated with the pre-averaging procedure, and we can expect discrepancies for the cyclization times due to the pre-averaging for small capture radius, as observed in Fig.  \ref{timeFig}. 
   

\section{Asymptotic behavior of the cyclization time}
\label{SectionScalings}

\subsection{Scalings of the MFCT in the Wilemski-Fixman theory}
\label{ScalingsWF}

Let us now derive the different scaling behaviors of the MFCT with $N$ and the capture radius $b$, first in the Wilemski-Fixman approximation. The MFCT are then computed with Eq. (\ref{ExpressionTMarkovianCentre}). 
We first consider the limit of small capture radius, $b\rightarrow0$ at fixed chain length $N$. If $b=0$, the integral (\ref{ExpressionTMarkovianCentre}) is  divergent due to the linear behavior of the MSD function $\psi$ at short times. Introducing the effective short time diffusion coefficient $D_s$ (such that $\psi\simeq_{t\rightarrow0}2D_st$), and considering that the integral (\ref{ExpressionTMarkovianCentre}) is governed by the short time regime, we get
\begin{align}
T&\simeq L^{3} \int_{0}^{\infty} dt \frac{e^{-b^{2}/4D_s t}}{(2D_s t )^{3/2}} = \frac{\sqrt{\pi} \ l_0^3\ (N-1)^{3/2}}{\sqrt{2}\ D_s\ b}. \label{SSS}
\end{align}
As discussed above, the actual value of $D_s$ could be quantitatively underestimated by the pre-averaging procedure, which computes it for equilibrium rather than looped configurations. Nevertheless, $D_s$ remains of the order of $k_BT/(6\pi\eta l_0)$. The scaling (\ref{SSS}) is similar to that appearing for free-draining flexible (Rouse) chains \cite{Amitai2012,Pastor,ThomasNat,ThomasCyclization}. It is known that such scaling does not depend on the structure of the chain since it is the same for semi-flexible chains \cite{ThomasSemiflex} and for hyperbranched structures; here it is clear that it also appears in the presence of hydrodynamic interactions.  
Eq. (\ref{SSS}) means that in the small capture radius regime, the MFCT is (up to a prefactor) the time needed for a diffusive particle to find a target of size $b$ in a confining volume $L^3$ \cite{Condamin2007,Benichou:2014fk}, and does not result from the collective dynamics of the monomers. 
  
\begin{figure}[h]
\begin{center}
\includegraphics[scale=0.45]{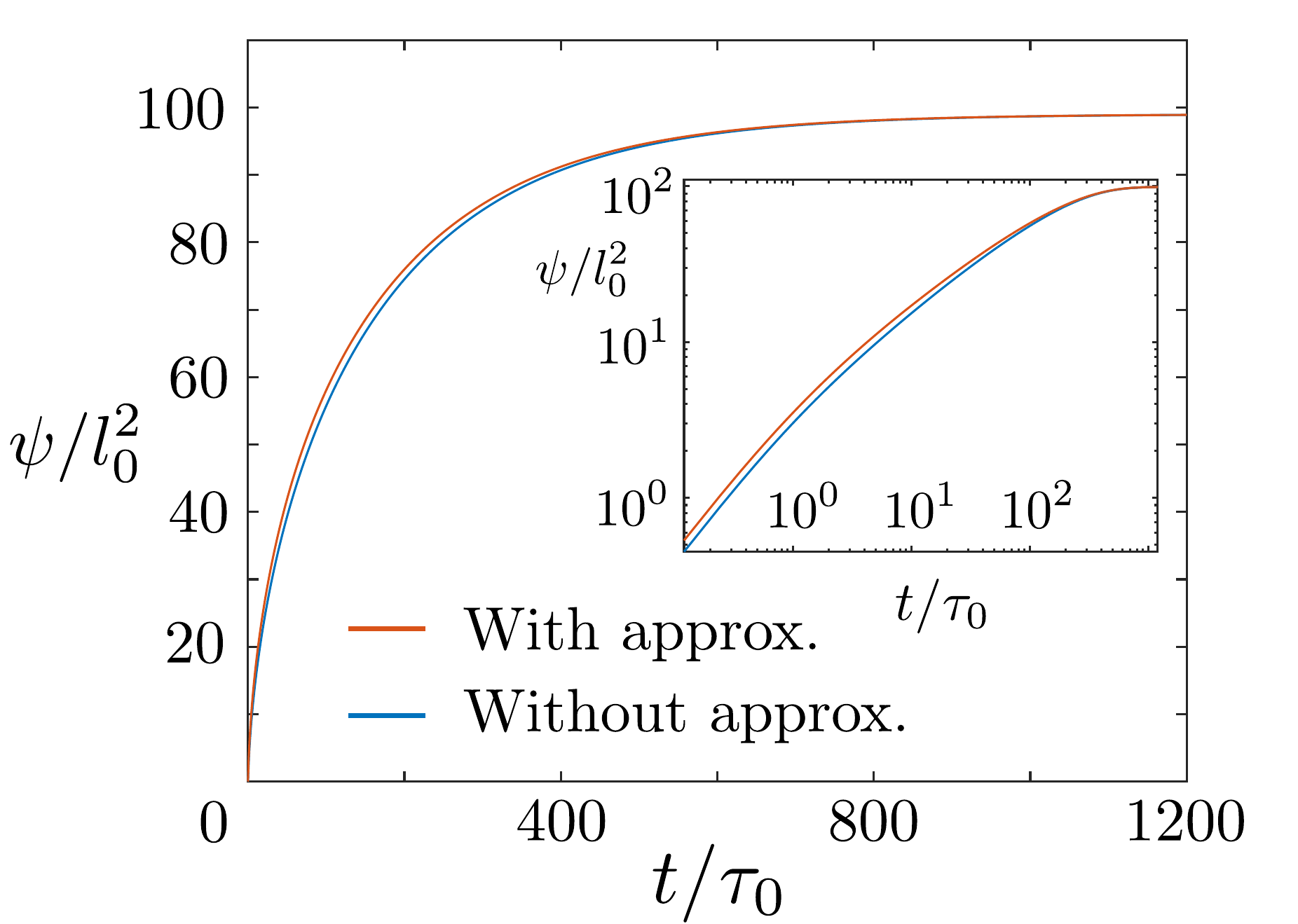} 
\end{center}
\caption{\textbf{Validity of the dominant diagonal approximation}. 
The red and blue curves show  the function $\psi(t)$ with and without the dominant diagonal approximation for $N=100$, respectively. Inset: same figure in double logarithmic scales.}
\label{CompPsi}
\end{figure}

We now consider the scaling of the MFCT in the limit of long chains. In this limit, we use the commonly used dominant diagonal approximation\cite{Edwards,Chak}, in which the orthogonal matrix $Q$ that diagonalizes the Laplacian matrix $M$ is assumed to diagonalize also the product $DM$. One can check in Fig. \ref{CompPsi} that the MSD function $\psi$ calculated with this approximation is very close to its exact value even for moderately large $N$. In this approximation, the correlation function in the continuous limit can be shown\cite{Edwards,Chak} to be
\begin{equation}
\phi(t)\simeq \sum_{p \; \text{odd}} \frac{8}{p^{2} \pi^{2}} e^{-p^{3/2}t/\tau_1} \label{Sum},
\end{equation}
where $\tau_1$ is the slowest relaxation time scale  of the chain,
\begin{align}
\tau_1=\frac{3 \eta l_0^3}{\sqrt{\pi} k_BT}N^{3/2}.\label{DefTau_1}
\end{align}
In the limit $t/\tau_1\ll 1$, the sum  (\ref{Sum}) can be replaced by a continuous integral, leading to the identification of the short time subdiffusive behavior
\begin{align}
\psi(t)= \langle [x_{\mathrm{ee}}(t)- x_{\mathrm{ee}}(0)]^2\rangle \simeq \frac{8\Gamma(1/3)}{3^{2/3}\pi^{5/3}}\left(\frac{k_BT\, t}{\eta}\right)^{2/3} \label{SubDIff}
\end{align}
which shows no dependence on the bond size $l_0$ or the chain length $N$. Here $\Gamma(\cdot)$ represents the Gamma function.  
Defining $\tau=t/\tau_1$, and introducing the rescaled functions $\phi(t)=\Phi(t/\tau_1)=\Phi(\tau)$ and $\psi(t)=l_0^2N\Psi(\tau)$, with $\Psi$ and $\Phi$ independent on $N$ in the continuous limit, 
a simple change of variable in Eq. (\ref{ExpressionTMarkovianCentre}) leads to
\begin{equation}
T=\tau_1 \int_{0}^{\infty} d\tau \left[ \frac{e^{-\tilde{b}^{2}\Phi(\tau)^{2}/2\Psi(\tau)}}{\Psi(\tau)^{3/2}} -1 \right]\equiv\tau_1 f(\tilde{b}) \label{8493}
\end{equation}
with $\tilde{b}=b/(l_{0}\,\sqrt{N})$ and $f$ a dimensionless function. Equation (\ref{8493}) finally gives the behavior of the mean cyclization time for continuous Zimm chains. 

We focus now on the asymptotics of $f$ for small rescaled capture radius $\tilde{b}$, which is highly dependent of the short time behavior of $\Psi\simeq \kappa\tau^{2/3}$, where $\kappa$ is deduced from Eqs. (\ref{DefTau_1},\ref{SubDIff}) and reads 
\begin{align}
\kappa=8\Gamma(1/3)/\pi^2. \label{DefKappa}
\end{align}
Because of this subdiffusive behavior, replacing $\tilde{b}$ by zero in (\ref{8493}) leads to a divergent integral, meaning that $f$ diverges with $\tilde{b}$ for small capture radii.  
The reason for this divergence is that, when $\tau\sim b^3$, the term $\tilde{b}^2/\Psi$ becomes of order $1$ and cannot be replaced by $0$. This suggests to introduce an intermediate "time" scale $\varepsilon$ such that $\tilde{b}^3\ll  \varepsilon\ll 1$, and to split the integral  (\ref{8493})  into two contributions, leading to
\begin{align}
f(\tilde{b})\simeq \int_0^{\frac{\epsilon}{\tilde{b}^3}}du\frac{e^{-\frac{1}{2\kappa u^{2/3}}}}{\kappa^{3/2}u}+
\int_{\epsilon}^{\infty}d\tau\left(\frac{1}{[1-\Phi^2]^{3/2}}-1\right) \label{993}
\end{align}
in which we have used the short time expression (\ref{SubDIff}) for $\tau<\varepsilon$, and simply set $\tilde{b}=0$ in the contribution coming from the large times. In the joint limit $\varepsilon\rightarrow0$ and $\varepsilon/\tilde{b}^3\rightarrow\infty$, the integral (\ref{993}) can be recast under the form 
\begin{align}
f(\tilde{b})&\simeq -\frac{\ln(\tilde{b}^3)}{\kappa^{3/2}}+ \int_0^{\infty}du \left[\frac{e^{-\frac{1}{2\kappa u^{2/3}}}}{\kappa ^{3/2}u}-\frac{\theta(u-1)}{\kappa^{3/2}u}\right]\nonumber\\
&+\int_{0}^{\infty}d\tau\left(\frac{1}{[1-\Phi(\tau)^2]^{3/2}}-1-\frac{\theta(1-\tau)}{\kappa^{3/2}\tau}\right)
\end{align}
with $\theta(\cdot)$  the Heaviside step function. All integrals appearing in this equation are convergent, and  their numerical evaluation leads to the scaling form  
\begin{equation}
f(\tilde{b})\simeq \frac{3}{\kappa^{3/2}}(-\ln\tilde{b}+0.721...) 
\end{equation}
In other words, we have identified the asymptotic scaling for the mean cyclization time of long Zimm chains
\begin{align}\label{eq29}
T\simeq \frac{9\pi^{5/2}}{[8 \Gamma(1/3)]^{3/2}}\frac{\eta \ l_0^3\ N^{3/2}}{k_BT}\ln\left(\frac{2.06 \ l_0\sqrt{N}}{b}\right).
\end{align}
We note that the weak logarithmic dependence on the size of the target is due to the fact that the motion of a monomer is a marginally compact process in this regime (the dimension of the walk satisfies $d_w=d=3$) \cite{Condamin2007,Benichou:2014fk}. The scaling $T\sim N^{3/2}\log N$ is in particular consistent with the renormalization group approach of \cite{Friedman1989,Chak}. The effect of hydrodynamic interactions is clearly visible, since this scaling of $T$ with $N$ is very different from the case of flexible chains without hydrodynamic interactions, where $T$ for long chains scales as $N^2$ and is independent of the capture  radius. Finally, let us note that a non-Markovian analysis reproduces the  scaling of the MFCT of Eq. (\ref{eq29}), with however a different numerical prefactor.

\section{Case of an external target}
\label{External}

The  formalism presented above can be adapted to the case of an external target, which models an inter-molecular reaction between one monomer at the end of the chain and a fixed target of radius $b$ in a confined volume $V$. This case has already been studied  for a Rouse chain \cite{ThomasTarget}, and we here briefly describe how the case of a Zimm chain can be treated. Assuming that the target is at the origin, the quantity of interest is $\bold{R}=\bold{R}_{1}$ if we assume that the reactive monomer is the first one. We present below the scaling of the mean reaction-time $T$ with the target, for a chain confined in a large volume $V$ in Wilemski-Fixman approximation. 
Eq.\eqref{basis} remains valid, and by integrating it over all configurations such that $\bold{x}_{1}=\ve[0]$, one gets :
\begin{align}
T P_{\text{stat}}&(\bold{x}_{1}=\ve[0])= \nonumber\\
&\int_{0}^{\infty} dt \left[ P(\bold{x}_{1}=\ve[0]|\pi,0)-P(\bold{x}_{1}=0|\text{ini},0) \right].
\end{align}  
The propagators appearing above are \emph{a priori} the propagators in a confined volume. In the limit of large volume, we argue that $P_{\text{stat}}(\bold{x}_{1}=0)$ is simply equal to $1/V$. The other propagators are well defined in infinite volume. The Wilemski-Fixman approximation then consists in assuming that the reactive conformations given by $\pi$ are simply the equilibrium configurations such that $|\bold{x}_{1}|=b$. Hence, one has~:
\begin{equation}\label{external}
\frac{T}{V}=\int_{0}^{\infty} dt \left[\frac{e^{-r^{2}/[2\psi(t)]}}{[2 \pi \psi(t)]^{3/2}} - \frac{e^{-R_{0}^{2}/[2\psi(t)]}}{[2 \pi \psi(t)]^{3/2}}\right]
\end{equation}
where $R_{0}$ is the initial position of the reactive monomer and $\psi(t)=\langle [x_1(t)-x_1(0)]^2\rangle$ the MSD of the reactive monomer, whose asymptotic behavior is  
\begin{equation}
\psi(t)\simeq \left\{ \begin{aligned}
&\alpha'\, t \hspace{1cm} &\text{for}&\hspace{1cm} t \ll \tau_0 \\
&\frac{4\Gamma(1/3)}{\pi^{5/3}}\left(\frac{k_BT t}{3\eta}\right)^{2/3} &\text{for}&\hspace{1cm} \tau_0 \ll t \ll \tau_1 \\
&2D_{\mathrm{cm}}\; t &\text{for}&\hspace{1cm} \tau_1\ll t
\end{aligned}
\right.
\end{equation}
where $D_{\mathrm{cm}}$ is the large time center-of-mass diffusion coefficient.
For large $N$,  the following  expressions are found \cite{Teraoka}~:
\begin{align}
&\alpha'=\frac{4}{3\sqrt{\pi}} \frac{k_{B} T}{\pi  \eta l_{0}}\\
&D_{\mathrm{cm}}=\frac{4\sqrt{2}}{9\pi^{3/2}}  \frac{k_{B} T}{\eta l_{0}\sqrt{N}}
\end{align}
The expression of the mean reaction time $T$ in the Wilemski-Fixman approximation can then be deduced from Eq. (\ref{external}) and yields in the limit of chains starting far from the target ($R_0\rightarrow\infty$)
\begin{equation}
T\sim \left\{ \begin{aligned}
&V /(4 \pi D_{\mathrm{cm}} b) &\;\; \text{for}& \;\; \sqrt{N} l_{0} \ll b \\
&\nu\frac{V\eta}{k_BT}  \log\left(\frac{\sqrt{N}l_0}{ b}\right) &\;\; \text{for} &\;\; l_{0} \ll b\ll \sqrt{N} l_{0}  \\
&V /(2 \pi \alpha'b) & \;\; \text{for} &\;\;b \ll l_{0}  \\
\end{aligned}
\right.\label{externalscaling}
\end{equation}
where $\nu$ is a numerical coefficient that reads
\begin{align}
\nu=\frac{9\pi}{8^{3/2}\Gamma(1/3)^{3/2}}.
\end{align}

As in the case of cyclization dynamics, the scaling of the mean reaction time with $N$ is different from the case of a  Rouse chain. In particular,  for intermediate capture radii $r$, the very weak (logarithmic) dependence on $N$ leads to reaction times that can be significantly shorter than for a Rouse polymer, for which\cite{ThomasTarget} $T\propto V \sqrt{N}$. The non-Markovian analysis of the reaction time with an external target closely follows the steps developed above for the determination of the MFCT and is not described here. As in the case of the cyclization dynamics, the scaling laws with $r$ and $N$ would be the same as predicted by the Wilemski-Fixman treatment (up to numerical prefactors), given in Eq. (\ref{externalscaling}).


\section{Conclusion}

In conclusion, we have presented in this paper a theoretical analysis of the closure kinetics of a polymer with hydrodynamic interactions. We have provided both a Markovian (Wilemski-Fixman)  and a non-Markovian analytical approach, based on a recent method introduced in \cite{ThomasNat,ThomasCyclization,ThomasTarget,Benichou:2015bh}. 
Although our theory relies on the preaveraging of the mobility tensor (Zimm dynamics), it is shown to reproduce very accurately the results of numerical simulations of the complete non linear dynamics. It is found that the Markovian treatment significantly overestimates cyclization times (up to a factor 2), showing the importance of memory effects in the dynamics. Such non-Markovian effects can be understood by analyzing the distribution of the polymer conformations at the instant of reaction, which is found to significantly depart from the equilibrium distribution. 

In addition, we derived asymptotic expressions of the mean cyclization time with the polymer size $N$ and capture radius $b$, which are identical in both Markovian and non-Markovian approaches, but with different prefactors. We computed the precise values of the prefactors in the case of the Wilemski-Fixman approach. 
In particular, it is found that the scaling of the MFCT for large $N$ is given by $T\sim N^{3/2}\ln (N/b^2)$, whereas for the Rouse chain one has $T\sim N^{2}$ (see Ref.~\cite{Benichou:2015bh} for review). Hydrodynamic interactions therefore change both the dependence on $N$ and $b$.  This difference comes from  the fact that subdiffusive exponent that characterize the monomer dynamics at intermediate length scales are different in both models. 

The present work demonstrates that the physics of cyclization kinetics in realistic models of polymers can be described by taking into account non-Markovian effects, which turn out to be much more important than the errors due to the approximate treatment of hydrodynamic interactions.

\acknowledgements
N.L., T.G., O.B., and R.V. acknowledge the support of the Campus France (project No. 28252XE) and of the European Research Council starting Grant No. FPTOpt-277998. M.D. and A.B. acknowledge the support of the DAAD through the PROCOPE program (project No. 55853833) and of the DFG through Grant No. Bl 142/11-1 and through IRTG Soft Matter Science (GRK 1642/1).  

\appendix 

\section{Pre-averaged Rotne-Prager tensor}
\label{appPreav}
We give in this section some details about the pre-averaging of the Rotne-Prage tensor, in order to quantify the differences obtained with the pre-averaged Oseen tensor. Let us first remind that the hydrodynamic interactions do not modify the equilibrium state of the polymer, which is that of a flexible Gaussian chain, hence the equilibrium probability density of the vector $\ve[r]_{ij}$ is 
\begin{align}
P_{\mathrm{eq}}(\ve[r]_{ij})=(2\pi\sigma_{ij})^{-3/2}e^{-\ve[r]_{ij}^2/(2\sigma_{ij})},
\end{align}
where $\sigma_{ij}=l_0^2\vert i-j\vert$ is the variance of each spatial coordinate of $\ve[r]_{ij}$ at equilibrium. The pre-averaged Rotne-Prager tensor is defined as
\begin{align}
\bar{\ve[D]}_{ij}=\int d\ve[r]_{ij}  \ P_{\mathrm{eq}}(\ve[r]_{ij}) \ve[D]_{ij}(\ve[r]_{ij}) \label{0593}.
\end{align}
We note that the average over all orientations of the non-isotropic tensor appearing in the expression of $\ve[D]_{ij}$ is
\begin{align}
\left \langle  \frac{ \ve[r]_{ij} \otimes  \ve[r]_{ij} }{r_{ij}^{2}} \right \rangle&= \frac{1}{3}\, \bold{I} 
\end{align}
Using this relation, the integral over rotational degrees of freedom can be performed in Eq.~(\ref{0593}); using Eqs.~(\ref{RP1}),(\ref{RP2}) we obtain an integral over radial components only:
\begin{align}
\bar{\ve[D]}_{ij}=k_BT \ \ve[I] \int_0^{\infty} d\rho  \  \frac{4\pi\rho^2e^{-\frac{\rho^2}{2\sigma_{ij}}}}{(2\pi\sigma_{ij})^{3/2}}
\Bigg[\frac{H(\rho-2a)}{6\pi\eta\rho}& \nonumber\\
+\frac{H(2a-\rho)}{6\pi\eta a}\left(1-\frac{\rho}{4 a}\right)& \Bigg].
\end{align}
with $H$ the Heaviside step function. The result of this integral in the limit of small $a$ is
\begin{align}
\bar{\ve[D]}_{ij}\simeq \frac{k_BT}{6 \pi \eta} \sqrt{\frac{2}{ \pi \sigma_{ij}}} \left(1-\frac{a^{2}}{3 \sigma_{ij}}\right) \, \bold{I}.
\end{align}
Note that for $a=0$ one recovers the pre-averaged Oseen tensor (\ref{PreaveragedTensor}). The corrections are of order $a^2/(3l_0^2\vert i-j\vert)$; with our choice of parameter $a=l_0/4$ these corrections are always less than $\simeq 2\%$, which justifies to chose to take the averaged Oseen tensor instead of the averaged Rotne-Prager tensor in the analysis.




\section{Explicit expressions of the dynamic quantities $\alpha_i,\beta_i,\phi,\psi$}
\label{appProp}
We consider the Fokker-Planck equation describing the chain dynamics with the pre-averaged mobility tensor. Since this tensor is isotropic, we consider the dynamics of a single spatial component, say $x_i$, of the vector positions $\ve[x]_i$. To the Langevin equation \eqref{Langevin} we can associate the Fokker-Planck equation
\begin{equation}
\frac{\partial P }{\partial t}= \displaystyle \sum_{i,j=1}^{N} \frac{\partial}{\partial x_{i}}  D_{ij} \left( \frac{\partial P }{\partial x_{j}} + \sum_{k=1}^N M_{jk}F_{k} P  \right).
\label{FP_eq}
\end{equation}
By replacing the mobility tensor by its pre-averaged version, Gaussian solutions do exist. The evolution of the mean vector $\mu=(\mu_1,...,\mu_N)$ and the covariance matrix $\Gamma_{ij}$ of $x_i,x_j$ with time can be found  by multiplying \eqref{FP_eq} by $x_{i}$ or $x_{m} x_{n}$, integrate over all $x_{k}$ and use the divergence theorem, resulting in\cite{VanKampen1992}
\begin{align}
&\dfrac{d \mu}{dt}= -K \mu \label{propagation}\\ 
&\dfrac{d  \Gamma }{dt}= 2D-K\Gamma - \Gamma \ ^tK 
\end{align}
where $K=D M$ is the product of the pre-averaged mobility matrix $D$ by the  connectivity matrix $M$. To solve these equations, it is convenient to introduce modes, which diagonalize $K$ (which ones coincide with Rouse modes if $D_{ij}\propto \delta_{ij}$). We define an invertible matrix $P$ such that $PKP^{-1}=\text{Diag} (\nu_{1},...,\nu_{N})$ is  diagonal, with $0=\nu_1<\nu_2<\nu_3<...$. The vanishing eigenvalue is associated to the translational motion of the chain, while all other eigenvalues describe the internal conformational degrees of freedom. We consider the amplitudes of the Zimm modes, defined by
\begin{align}
a_i=\sum_{j=1}^N P_{ij}x_j
\end{align}
and the associated mean vector $\mu_i=\langle a_i\rangle$ and covariance matrix  $\kappa_{ij}=\mathrm{Cov}(a_i,a_j)$ read 
\begin{align}
	u=P\mu ; \hspace{0.3cm} \kappa=P  \  \Gamma \ {}^tP; \hspace{0.3cm} B=2P\ D  \ ^tP
\end{align}
Note that $P$ is not an orthogonal matrix. The evolution of $u$ and $\kappa$ reads
\begin{align}
&\dfrac{d \,u_{i}}{dt}= -\nu_{i} \,u_{i}\\
&\dfrac{d \kappa_{ij}}{dt}= -(\nu_{i}+\nu_{j}) \,\kappa_{ij} + B_{ij}\label{DynamicsCovModes}
\end{align}
The solutions are~:
\begin{align}
u_{i} (t)&= u_{i}(0) \,e^{-\nu_{i}  t} \\
\kappa_{ij} (t)&=\left(\kappa_{ij}(0)-\frac{B_{ij}}{\nu_{i}+\nu_{j}}\right)\,e^{-(\nu_{i}+\nu_{j})\,t}+\frac{B_{ij}}{\nu_{i}+\nu_{j}}
\label{prop}
\end{align}
At large times, the covariance matrix reaches the stationary value (for $i,j\ge2$):
\begin{equation}
	\kappa_{ij}^{s}=\frac{B_{ij}}{\nu_{i}+\nu_{j}}
\end{equation}
We introduce the set of coefficients $\tilde{c}$, 
\begin{align}
	\tilde{c}_{k}=(P^{-1})_{kN}-(P^{-1})_{k1}=P^{-1}h
\end{align}
where we remind that $h=(-1,0,....,0,1)^t$. The vector $\tilde{c}$ is such that
\begin{align}
x_{\mathrm{ee}}=\sum_{i=1}^N \tilde{c}_i a_i
\end{align}
Note that $\tilde{c}_1=0$, meaning that the motion of the drift center is not involved in the evolution of the internal variable $x_{\mathrm{ee}}$. 
The equilibrium end-to-end distance is~:
\begin{equation}
L^{2}=\lim_{t\rightarrow\infty}  {}^t h\Gamma h=\sum_{m,n= 2}^N \tilde{c}_{m} \,\kappa^{s}_{mn}  \,\tilde{c}_{n} 
\end{equation}
Similarly, we obtain for $\alpha_{i}$
\begin{align}
\alpha_{i}=\sum_{j=2}^{N}\sum_{k=1}^{N} (P^{-1})_{ij} \, \kappa_{jk}^{s} \, \tilde{c}_{k} 
\end{align}
Consider the correlation matrix $C_{ij}(t)=\langle a_i(t)a_j(0)\rangle$, starting from an equilibrium configuration. This matrix is a solution of 
\begin{align}
	\partial_t C_{ij}=-\nu_i C_{ij} \ ; \ C_{ij}(0)=\kappa_{ij}^s
\end{align}
The solution is straightforward: $C_{ij}=\kappa_{ij}^s e^{-\nu_i t}$. Given that $\langle x_{\mathrm{ee}}(t)x_{\mathrm{ee}}(0)\rangle=\sum_{ij}\tilde{c}_i\tilde{c}_jC_{ij}$, we obtain for the function $\phi$  [defined in Eq.~(\ref{DefPhi})]
\begin{equation}
\phi(t)=\frac{1}{L^2}\sum_{m,n=2}^N  \tilde{c}_{n} \, \kappa_{mn}^{s} \, \tilde{c}_{m} e^{-\nu_{m} t} \label{ExpressionPhi}
\end{equation}
Next, we denote $\kappa^*$ the covariance matrix of $a_i,a_j$ at equilibrium with the constraint of fixed $x_{\mathrm{ee}}$, which from (\ref{ConditionalCovariance}) reads
\begin{align}
	\kappa_{ij}^*=
\kappa_{ij}^{s}-\sum_{m,n=1}^{N}\frac{\kappa_{im}^s \kappa_{jn}^s  \tilde{c}_m\tilde{c}_n}{L^2}  \label{CovStatStar}
\end{align} 
Consider now
\begin{align}
\kappa_{ij}(t)=\mathrm{Cov}(a_i(t),a_j(t)\vert x_{\mathrm{ee}}(0)=0)
\end{align}
which is related to $\beta_i$ by
\begin{align}
\beta_i=\sum_{jk=1}^N(P^{-1})_{ij}\kappa_{jk} \tilde{c}_k \label{0594}
\end{align}
Taking the covariance matrix (\ref{CovStatStar}) as an initial condition for the dynamics (\ref{DynamicsCovModes}), we obtain 
\begin{equation}
\kappa_{ij}(t)=\kappa_{ij}^{s}-\frac{\sum_{n}\kappa_{in}^{s} \tilde{b}_{n} \sum_{m}\kappa_{jm}^{s} \tilde{b}_{m}}{L^{2}} e^{-(\nu_{i}+\nu_{j})t}
\end{equation}
Using (\ref{ExpressionPhi}) and (\ref{0594}), we obtain for $\beta_i$
\begin{align}
&\beta_{i}=\alpha_{i}-\phi(t) \sum_{j} \sum_{k} P^{^{-1}}_{ij} \, \kappa_{jk}^{s} \, e^{-\nu_{j} t}\, \tilde{b}_{k}
\end{align}
Finally, the function $\psi=\beta_N-\beta_1$ reads 
\begin{equation}
\psi(t)= L^{2}(1-\phi^{2}(t))
\end{equation}
where we have again used  (\ref{ExpressionPhi}). Hence, the expressions of all the dynamical quantities $\alpha_i,\beta_i,\phi,\psi$ are given explicitly in this section.

\section{Derivation of the self-consistent equations \eqref{mom}}
\label{appAutoCoh}
We present here a derivation of the set of equations \eqref{mom} for the moments $m_i^{\pi}$, which adapts the method used for Rouse polymers\cite{ThomasCyclization}. First, we multiply \eqref{basis} by $x_{iz }\delta(\ve[x]_N-\ve[x]_1-R_f\hat{\ve[e]}_z)$ (where $\hat{\ve[e]}_z$ is the unit vector in the vertical direction, and $R_f$ is fixed and satisfies $0<R_f<b$) and integrate over all configurations, to get~:
\begin{align}
&T p_{\text{stat}} ( R_{f}\hat{\ve[e]}_z) \mathbb{E}_{\text{stat}}(x_{i z}\vert  R_{f})=\nonumber\\
&\int_{0}^{\infty} dt \int d\Omega [  p(R_{f}\hat{\ve[e]}_z,t | \pi_{\Omega}, 0) \mathbb{E}(x_{i z},t| R_{f}\hat{\ve[e]}_z,t; \pi_{\Omega},0) \nonumber\\
&  - p(R_{f}\hat{\ve[e]}_z,t | P_{\text{ini},\Omega}, 0) \mathbb{E}(x_{i z},t| R_{f}\hat{\ve[e]}_z,t;P_{\text{ini},\Omega},0 )] \label{94593}
\end{align}
with $p$ representing the probability distribution function of the end-to-end vector, $p(R_{f}\hat{\ve[e]}_z,t | \pi_{\Omega}, 0)$ is the probability that $\ve[r]_{\mathrm{ee}}=R_f\hat{\ve[e]}_z$ at $t$ with an initial distribution $\pi_{\Omega}$, $\mathbb{E}(x_{i z},t| R_{f}\hat{\ve[e]}_z,t; \pi_{\Omega},0) $ is the conditional average of the $z$ coordinate of $\ve[x]_i$  at $t$ given that $\ve[r]_{\mathrm{ee}}=R_{f}\hat{\ve[e]}_z$ at the same time $t$ and starting from $\pi_{\Omega}$ initially; other notations are similar. The mention ``$P_{\text{ini},\Omega}$" means the equilibrium distribution with 
$\ve[r]_{\mathrm{ee}}=R_{0} \bold{u}_{r}(\Omega)$ initially, we perform the average over initial end-to-end distances at the end of the calculation.

Let us remind the following formula for the conditional mean of a Gaussian variable $X$ given that a second Gaussian variable $Y$ takes the value $Y_0$: 
\begin{align}
	 \mathbb{E}(X\vert Y=Y_0)= \mathbb{E}(X)-\frac{\mathrm{Cov}(X,Y)}{\mathrm{Cov}(Y,Y)}[ \mathbb{E}(Y)-Y_0].	\label{ConditionalMean} 
\end{align}
A similar general formula for the conditional covariances of Gaussian variables is
\begin{align}
	 \mathrm{Cov}(X_1&,X_2\vert Y=Y_0)=\nonumber\\
 &\mathrm{Cov}(X_1,X_2)-\frac{\mathrm{Cov}(X_1,Y)\mathrm{Cov}(X_2,Y)}{\mathrm{Cov}(Y,Y)}.	\label{ConditionalCovariance} 
\end{align}
Now, we consider a fixed angular direction $\Omega$ and define $\theta$ the angle with the vertical direction. We write 
\begin{align}
x_{i z}=x_{i r} \cos \theta - x_{i \theta} \sin \theta,  \label{90583}
\end{align}
with $x_{ir}$ the component of $\ve[x]_i$ in the direction $\hat{\ve[u]}(\Omega)$. We note that conditioning the end-to-end vector to have the value $R_f\hat{\ve[e]}_z$  imposes that its component in the direction $\hat{\ve[u]}(\Omega)$ takes the value $R_f\cos\theta$. Then, applying (\ref{ConditionalMean}) and using the definitions of $R_{\pi},\beta_i,\psi$, we get:
\begin{align}
 \mathbb{E}(x_{i r},t| R_{f}\hat{\ve[e]}_z,t, \pi_{\Omega},0)&=\mu_{i}^{\pi}-\frac{\beta_i}{\psi} (R_{\pi} - R_{f} \cos \theta).
\end{align}
Applying the same reasoning in the direction normal to $\ve[u](\Omega)$, we get
\begin{align}
 \mathbb{E}(x_{i \theta},t| R_{f}\hat{\ve[e]}_z,t, \pi_{\Omega},0)&=-\frac{\beta_i}{\psi}  R_{f} \sin \theta.
\end{align}
Inserting these expressions into (\ref{90583}) leads to
\begin{align}
 \mathbb{E}(x_{i z},t| R_{f}\hat{\ve[e]}_z,t, \pi_{\Omega},0)=&\cos \theta \left(\mu_{i}^{\pi}-\frac{\beta_i}{\psi}R_{\pi}\right) +\frac{R_f\beta_i}{\psi}.  \label{9543}
\end{align}
Let us pose now 
\begin{align}
&\mu_i^{\mathrm{stat},R_0}=\mathbb{E}(x_{i z}(t)| x_{\mathrm{ee}}(0)=R_{0},0),\label{DefMuISTat}\\
&R=\mathbb{E}(x_{\mathrm{ee}}(t)| x_{\mathrm{ee}}(0)=R_{0},0),
\end{align}
where initial equilibrium conditions  (apart from the constraint for  $x_{\mathrm{ee}}(0)$) are understood. A reasoning similar to that leading to Eq.~(\ref{9543}) gives
\begin{align}
\mathbb{E}(x_{i z},t| R_{f}\hat{\ve[e]}_z,t; &P_{\text{ini},\Omega},0)=\nonumber\\
&\cos \theta \left(\mu_{i}^{\mathrm{stat},R_0}-\frac{\beta_i}{\psi}R\right) +\frac{R_f\beta_i}{\psi}
\end{align}
Similarly, from (\ref{ConditionalMean}) and (\ref{DefAlpha_i}), we obtain
\begin{align}
\mathbb{E}_{\text{stat}}(x_{i z} \vert R_{f}\hat{\ve[e]}_z)=&\frac{\alpha_i }{L^2 }  R_{f}
\end{align}

Finally, the propagators for the end-to-end distance read~:
\begin{align}
&p(R_{f}\hat{\ve[e]}_z,t| P_{\text{ini},\Omega},0)=\frac{1}{(2\pi\psi)^{3/2}} e^{-\frac{(R_{f} \bold{u}_{z}-R \bold{u}_{r})^{2}}{2 \psi}}, \\
&p(R_{f}\hat{\ve[e]}_z,t | \pi_{\Omega},0)=\frac{1}{(2\pi\psi)^{3/2}} e^{-\frac{(R_{f} \bold{u}_{z}-R_{\pi} \bold{u}_{r})^{2}}{2 \psi}} ,\\
&p_{\text{stat}}(R_{f}\hat{\ve[e]}_z)=\frac{1}{(2\pi L^{2})^{3/2}} \exp{\left(-\frac{R_{f}^{2}}{2 L^{2}}\right)}.
\end{align}
All the terms appearing in (\ref{94593}) have been evaluated, the self-consistent equation becomes
\begin{align}
\int_{0}^{\infty} dt \int_{0}^{\pi}  d\theta &\frac{\sin \theta}{2} \left[\frac{1}{(2\pi\psi)^{3/2}} e^{-\frac{(R_{f} \bold{u}_{z}-R \bold{u}_{r})^{2}}{2 \psi}}\right.  \nonumber\\
&\times \left(\cos \theta \,\mu_{i}^{\pi}(t)-\frac{\beta_{i}}{\psi} (\cos \theta \, R_{\pi}  - R_{f})\right) \nonumber \\
&- \frac{1}{(2\pi\psi)^{3/2}}e^{-\frac{(R_{f} \bold{u}_{z}-R_{\pi} \bold{u}_{r})^{2}}{2 \psi}}  \nonumber\\
& \times \left. \left(\cos \theta \,\mu_{i}^{\text{stat},R_{0}}(t)-\frac{\beta_{i}}{\psi} (\cos \theta \,R - R_{f})\right) \right]  \nonumber\\
&=T \,R_{f} \frac{\alpha_{i}}{L^{2}} p_{\text{stat}}(R_{f}).
\end{align}
This equation should be verified for any $R_{f}$ between $0$ and $b$. We choose to write it in the limit $R_f\rightarrow0$ (more precisely, we develop both expressions at first order in $R_f$). Noting that $\hat{\ve[u]}_r\cdot\hat{\ve[e]}_z=\cos\theta$, the integration over $\theta$ can be performed, leading to
\begin{align}
\int_{0}^{\infty} dt &\left\{\frac{e^{-R_{\pi}^{2}/2 \psi}}{(2\pi\psi)^{3/2}} \left[\frac{\beta_{i}}{\psi}+\frac{R_{\pi}}{3 \psi}(\mu_{i}^{\pi}-\frac{\beta_{i}}{\psi} R_{\pi})\right]\right. \nonumber \\
&\left. -\frac{e^{-R^{2}/2 \psi}}{(2\pi\psi)^{3/2}} \left[\frac{\beta_{i}}{\psi}+\frac{R}{3 \psi}(\mu_{i}^{\text{stat},R_{0}}-\frac{\beta_{i}}{\psi} R)\right]\right\} \nonumber \\
&=\;T\,\frac{\alpha_{i}}{L^{2}} \frac{1}{(2 \pi L^{2})^{3/2}}. \label{8472}
\end{align}
We can eliminate $T$ by replacing its expression~:
\begin{equation}
\frac{T}{ L^{3}}=\int_{0}^{\infty} \frac{dt}{\psi^{3/2}} \left\{ e^{-R_{\pi}^{2}/2 \psi}-e^{-R^{2}/2 \psi} \right\}.
\end{equation}
Let us now find the dependence of $R$ and $\mu_{i}^{\text{stat},R_{0}}$ with $R_0$. 
From (\ref{DefMuISTat}) and  (\ref{ConditionalCovariance}), we deduce 
\begin{align}
\mu_i^{\mathrm{stat},R_0}=R_0\frac{\mathrm{Cov}(x_i(t),x_{\mathrm{ee}}(0))}{L^2}
\end{align}
Furthermore, applying the formula for the conditional covariances in  (\ref{DefBeta_i}), we get
\begin{align}
\beta_i= \ & \mathrm{Cov}(x_i(t)x_{\mathrm{ee}}(t))\nonumber\\
&-\frac{\mathrm{Cov}(x_i(t)x_{\mathrm{ee}}(0))\mathrm{Cov}(x_{\mathrm{ee}}(t)x_{\mathrm{ee}}(0))}{\mathrm{Cov}(x_{\mathrm{ee}}(0)x_{\mathrm{ee}}(0))}
\end{align}
From which we obtain 
\begin{align}
\mu_{i}^{\text{stat},R_{0}}=\frac{R_0(\alpha_i-\beta_i)}{L^2\phi(t)}, \label{mu_istatR0}
\end{align} 
and 
\begin{align}
R=\mu_{N}^{\text{stat},R_{0}}-\mu_{1}^{\text{stat},R_{0}}= R_0\phi(t)  \label{RR_0}
\end{align} 
The last step consists in adapting the reasoning to the case of initial end-to-end distance that is distributed $p_r(R_0)=R_0^2e^{-R_0^2/2L^2}/\int_b^{\infty}r^2 e^{-r^2/2L^2}dr$. This can be achieved by keeping an average over $R_0$ at all steps of the derivation. Averaging (\ref{8472}) with respect to $R_0$  [by using (\ref{mu_istatR0},\ref{RR_0})] leads to the final form of the self-consistent equation \eqref{mom} for the moments given in the main text.

\section{Numerical methods}
\label{num}
We performed the numerical integration of the stochastic equation (\ref{Langevin}) by using either the full Rotne-Prager tensor or its pre-averaged form. At each time step of size $\Delta t$, the positions of the monomers evolve according to the algorithm of Ermak and McCammon\cite{BrownianDynamics}:
\begin{equation}
	\mathbf{x}_{i}(t+\Delta t)=\mathbf{x}_{i}(t) + \sum_{j=1}^N \frac{\mathbf{D}_{ij}(t) \mathbf{F}_{j}(t)}{k_{B}T} \Delta t + \boldsymbol{\xi}_{i}(\Delta t) \label{EvolutionPositions}
\end{equation}
with the same notations as before. $\boldsymbol{\xi}_{i}(\Delta t)$ is a random Gaussian noise with zero mean and covariance $\left \langle \boldsymbol{\xi}_{i}(\Delta t) \boldsymbol{\xi}_{j}(\Delta t) \right \rangle = 2 \mathbf{D}_{ij}(t) \Delta t$.
The generation of the $3N$ random numbers $\boldsymbol{\xi}_{i}$ requires to find a Cholesky decomposition of the mobility tensor, which can be done in $N^{3}$ operations.
In the case of simulations using the pre-averaged mobility tensor (\ref{PreaveragedTensor}), this decomposition needs to be performed only once, whereas simulations using the full Rotne-Prager tensor require to perform a Cholesky decomposition at each time step, resulting in considerably longer computational times which prevented us to explore the same range of parameters. 
Initial configurations are generated from the equilibrium Gibbs-Boltzmann Gaussian distribution $P_{\mathrm{eq}}(\{\ve[x]\})\propto e^{-k \sum_{i=1}^{N-1}(\ve[x]_{i+1}-\ve[x]_{i})^2/2k_BT}$, and are rejected if the condition $\vert\ve[x]_N-\ve[x]_1\vert>b$ is not satisfied. Once an equilibrium configuration is generated, the positions evolve through (\ref{EvolutionPositions}) until reaching a configuration $\vert\ve[x]_N-\ve[x]_1\vert\le b$, the cyclization time $t_i$ for this run is recorded. The MFCT is finally found by ensemble averaging $t_i$ over many runs. The time step is chosen as suggested by Pastor \emph{et al.} \cite{Pastor}. Noting $R$ the end-to-end distance  :
\begin{equation}
\Delta t = \Delta _{\text{low}}+\Delta _{\text{high}} \sin \left(\frac{\pi}{6}(R^2-b^2)\right)
\end{equation}
if $R^2<b^2+3 l_{0}$, and
\begin{equation}
\Delta t = \Delta _{\text{low}}+\Delta _{\text{high}} 
\end{equation}
if not.

We choosed $\Delta_{\text{high}}=\sqrt{N} 10^{-5}$ and $\Delta_{\text{low}}=10^{-5}$, and we controlled that convergence was reached. 
The results of simulations are represented on Fig.~\ref{timeFig}. 

\bibliographystyle{apsrev}

\bibliography{biblioZimm}

\end{document}